\documentclass[doublecol]{epl2} 

\usepackage{graphicx}
\usepackage{color}
\usepackage{amsmath}
\usepackage{amssymb}
\usepackage[utf8]{inputenc}
\usepackage{comment}
\usepackage{dcolumn}
\usepackage{bm}
\usepackage{hyperref}
\usepackage{soul}
\usepackage{footmisc}
\usepackage{xfrac}
\usepackage{upgreek}

\newcommand{\Rmnum}[1]{\expandafter\@slowromancap\romannumeral  #1@}

\usepackage{bbold}

\title{On spectral properties of Generalized Kadanoff--Baym Ansatz}
\shorttitle{Title} 

\author{Miroslav Hopjan\inst{1}}
\shortauthor{Miroslav Hopjan}

\institute{                    
  \inst{1} Institute of Theoretical Physics, Wrocław University of Science and Technology, 50-370 Wrocław, Poland\\
}

\abstract{The Generalized Kadanoff–Baym Ansatz with mean-field propagators is increasingly used to simulate non-equilibrium quantum fermionic and bosonic systems. Compared with the full Kadanoff–Baym equations, its mean-field propagators substantially reduce computational cost, enabling simulations with computational effort that scales linearly with propagation time. However, the time-diagonal structure of the Generalized Kadanoff–Baym Ansatz obscures spectral properties within the collision integral of the transport equation. Here, we recover and investigate these hidden spectral properties using the extended Generalized Kadanoff–Baym Ansatz. For a Hubbard cluster, we compare ground-state spectral functions obtained with the extended Generalized Kadanoff–Baym Ansatz against those from standard Kadanoff–Baym equations. At moderate interaction strengths, the spectral functions show good agreement. At large interaction strengths, however, significant deviations emerge, and the spectral function obtained with the extended Generalized Kadanoff–Baym Ansatz becomes negative. These results demonstrate that, despite its computational advantages, the Generalized Kadanoff–Baym Ansatz has important limitations in strongly interacting regimes and may produce unphysical spectral properties when interactions become sufficiently large.}

\begin{document}

\maketitle

\section{Introduction}

The field theoretical approach to equilibrium and non-equilibrium systems in condensed matter$~$\cite{Martin59,Baym61,Baym62,Kadanoff62,Abrikosov63,Keldysh65,Hedin65,Langreth72,Danielewicz84,Meir92,Aryasetiawan98,Hedin99} was devised  
as an efficient description that avoids the exponential growth of the computational costs of the many-body problem with increasing system-size. The basic quantity of the approach is the single-particle Greens function, i.e., the spatial and temporal correlation function $G_{i,j}(t,t')$, where $i,j$ stands for the spatial coordinates and $t,t'$ for the time coordinates, which provide access to particle densities, particle currents, and dynamical properties via spectral functions. In the general case of time-dependent non-equilibrium systems, the equations of motion for the Greens functions are given by the Kadanoff--Baym equations (KBE) defined on the Martin-Schwinger-Keldysh contour. The numerical implementations of the KBE, both in closed and open quantum systems, have been the subject of the last three decades$~$\cite{Bonitz98,Kwong00,Dahlen07,BALZER08,
 Myohanen08,Myohanen09,Stan09,Friesen09,Friesen10a,Friesen10b,
Stefanucci13,Balzer13,Pavlyukh13,Hopjan14,Perfetto15,Sakkinen15,
Schuler16,Hopjan16,Lynn16,Schlunzen16a,Schlunzen16b,Bostrom16,Gullo16,Pavlyukh17,Schlunzen17a,Karlsson18a,Covito18,Pavlyukh18,Talarico19,Settino20,
Joost21,Ridley22,Pavlyukh23,Reeves24,Freericks06,Freericks07,Freericks09,Moritz10,Sentef13,Kemper13,Kemper14,Aoki14,Schuler20,
Kaye21,Dong22,Kaye23,Schroedter23,Schroedter24a,Schroedter24b} and relatively large systems have been reached compared to the system sizes available for exact numerical solutions. However, even the KBE becomes intractable for mesoscopic system sizes, as the numerical costs scale cubically, i.e., $N_t^3$, with the number $N_t$ of discrete time steps. Therefore, in recent years, there has been a tendency to reduce the costs of KBE solutions by, e.g., memory compression\cite{Kaye21,Kaye23,Środa25} or other solutions$~$\cite{Reeves24,Schroedter23,Schroedter24a,Schroedter24b}. 

The alternative way to reduce the numerical costs, which was introduced 40 years ago in Ref.$~$\cite{Lipavsky86}, is to compute only the time-diagonal part of the Greens function, i.e., the single-particle (or one-particle) density matrix $\rho_{i,j}(t)=G_{i,j}(t,t)$. Its time evolution is governed by the time-diagonal transport equation, which can be derived from the KBE by reduction to the time diagonal. The reduction of computational costs is based on the truncation of the reconstruction equations, known as the Generalized Kadanoff--Baym Ansatz (GKBA)$~$\cite{Lipavsky86}, together with the simplest mean-field Hartree-Fock approximation for the auxiliary retarded Greens function$~$\cite{Hopjan18}.
For this approximation, dubbed HF-GKBA, the reduction of the numerical costs by an apparent linear factor $N_t$ was reported in many studies 
\cite{Spicka05a,Spicka05b,Spicka05c,Velicky06,Velicky07,Velicky08,
Velicky10a,Spicka10,Velicky10b,Spicka12,Hermanns12,Spicka14,Kalvova14,Latini14,HERMANNS14,BarLev16,Spicka17,Schlunzen17b,Hopjan17,Perfetto18a,Perfetto18b,VinasBostrom18,Karlsson18b,Hopjan18,Kalvova18,Kalvova19,Perfetto19,VinasBostrom19,Perfetto20,Cosco20,Joost20,Schlunzen20,Maansson21,Pavlyukh21,Karlsson21,Spicka21,Meirinhos22,Pavlyukh22a,Perfetto22,Pavlyukh22b,Pavlyukh22c,Joost22,VinasBostrom22,Perfetto23,Tuovinen23,Kalvova23,Balzer23,Bonitz23,Reeves23a,Reeves23b,Cosco24,Kalvova24,Ostber24,Gopalakrishna24}. However, an even more dramatic speedup, by a factor $N_t^2$, can be achieved when the HF-GKBA integro-differential equations are reformulated as a set of differential equations, a scheme dubbed the G$_1$-G$_2$ method$~$\cite{Joost20,Schlunzen20}. Despite the success in reducing the computational costs\cite{Karlsson21,Balzer23,Bonitz23,Tuovinen23}, the HF-GKBA, and the mean-field-GKBA in general, are known to exhibit instabilities$~$\cite{Schlunzen20} and negative occupations$~$\cite{Schlunzen20,Pavlyukh22b}. These features of the HF-GKBA call for a better understanding of its spectral properties. 

Here, we aim to uncover the spectral properties of the underlying Greens function that are hidden in the HF-GKBA transport equation. To this end, we consider the extended GKBA (eGKBA)$~$\cite{Hopjan18} within the same HF approximation of the auxiliary retarded Greens function, dubbed HF-eGKBA. While the eGKBA was proposed in Ref.$~$\cite{Hopjan18}, it has not yet been numerically implemented. Thus, in this work, we numerically obtain the solution of the eGKBA and study its properties. As a model example, we study the equilibrium spectral functions of the six-site Hubbard model and compare the solutions of the KBE and HF-eGKBA. We find that, for moderate interaction strength, the spectral functions are in good agreement. However, for large interaction strength, the spectral functions start to differ, and we show that the HF-eGKBA can lead to a non-positive spectral function. The emergence of the negative spectral function in HF-eGKBA constitutes the main result of the present work and may lay the groundwork for understanding issues that can arise in the HF-GKBA time scheme, such as the emergence of negative densities and instabilities. 

The plan of the paper is as follows. We first review the non-equilibrium formulation of KBE, GKBA, and define the HF-GKBA flavor. We then discuss the non-equilibrium formulation of eGKBA and its HF-eGKBA flavor. After that, we turn to the formulation of the KBE and the HF-eGKBA equations for equilibrium homogeneous systems and present their numerical solution for the six-site Hubbard model. We conclude with the discussion and the possible implications of our work.

\section{Kadanoff--Baym equations}
The time ordered single particle (or one-particle) Greens function $G(t,t')$ is propagated according to the integro-differential Kadanoff--Baym equations$~$\cite{Baym61,Baym62,Kadanoff62,Keldysh65,Bonitz98,Stefanucci13,Balzer13,Hopjan14}, where the integration is done along the Martin--Schwinger--Keldysh contour $\gamma$. We note that we simplify the notation
by dropping the spatial indices; thus, $G(t,t')$ is to be understood as a matrix in the spatial coordinates. Breaking the contour integration, using the Langreth--Wilkins rules$~$\cite{Langreth72,Stefanucci13,Balzer13}, into real time integrals, the equations of motion read as
\begin{equation}\label{kbeeq_left}
\biggl[i\partial_t-h_{\rm HF}(t)\biggr]G^{\lessgtr}(t,t')=I^{\lessgtr}(t,t'),
\end{equation}
\begin{equation}\label{kbeeq_right}
-G^{\lessgtr}(t,t')\biggl[-i\overleftarrow{\partial_{t'}}-h_{\rm HF}(t')\biggr]=\bigl[I^{\lessgtr}(t',t)\bigl]^{\dagger}.
\end{equation}
Here, the symbols $I^{<,>}$ stand for the collision integrals
\begin{equation}\label{collision_left}
I^{<}(t,t')=\int^{\infty}_{-\infty}d\bar{t}[{\Sigma}^{<}(t,\bar{t})G^{A}(\bar{t},t')+{\Sigma}^{R}(t,\bar{t})G^{<}(\bar{t},t')],\\
\end{equation}
\begin{equation}\label{collision_right}
I^{>}(t,t')=\int^{\infty}_{-\infty}d\bar{t}[G^{>}(t,\bar{t}){\Sigma}^{A}(\bar{t},t')+G^{R}(t,\bar{t}){\Sigma}^{>}(\bar{t},t')], \\
\end{equation}
and the superscripts $<,>,A,R$ define lesser, greater, advanced, and retarded components, respectively, of the contour ordered Greens function $G$ and the self-energy ${\Sigma}$. The self-energy ${\Sigma}$ consists of an embedding part ${\Sigma}_{\rm emb.}$ that describes the effects of the environment and a correlation part ${\Sigma}_{\rm corr.}$ that takes into account the effects of interactions. Here, we consider only closed systems, and we thus set ${\Sigma}_{\rm emb.}=0$.

The equations of motion for the retarded Greens function read as
\begin{equation}\label{retarded_left}
\biggl[i\partial_t-h_{\rm HF}(t)\biggr]G^{R}(t,t')=\delta(t,t')+\int_{}d\bar{t}{\Sigma}^{R}(t,\bar{t})G^{R}(\bar{t},t'),
\end{equation}
\begin{equation}\label{retarded_right}
G^{R}(t,t')\biggl[i\overleftarrow{\partial_{t'}}-h_{\rm HF}(t')\biggr]=\delta(t,t')+\int_{}d\bar{t}G^{R}(t',\bar{t}){\Sigma}^{R}(\bar{t},t),
\end{equation}
and, similar equations, where the superscript $R$ is replaced by the superscript $A$, then apply to the advanced Greens function ${G}^{A}$. The Greens function components evolved by Eqs.$~$\eqref{kbeeq_left}-\eqref{retarded_right} satisfy the fundamental spectral identity
\begin{eqnarray}\label{spectral_identity}
{G}^{>}(t,t')-{G}^{<}(t,t')={G}^{R}(t,t')-{G}^{A}(t,t').
\end{eqnarray}
\section{Time-diagonal transport equation and reconstruction equations}
The KBE for lesser and greater components, Eq.$~$\eqref{kbeeq_left} and Eq.$~$\eqref{kbeeq_right}, can be alternatively formulated in terms of the equation of motion of the single particle density matrix $\rho(t)=-iG^{<}(t,t)$ and the so-called reconstruction equations for $G^{\lessgtr}(t,t')$. The exact equation for $\rho(t)$ can be derived from the difference between Eq.$~$\eqref{kbeeq_left} and its adjoint Eq.$~$\eqref{kbeeq_right}. This results in a transport equation on the time diagonal
\begin{equation}\label{transport}
\begin{split}
\partial_{t}\rho(t)+i[h_{HF}(t),\rho(t)]=-[I^{<}(t,t)+{\rm H.c.}],
\end{split}
\end{equation}
where $[...,...]$ stands for the commutator. This is, however, not a closed equation for $\rho(t)$, since Greens functions with off-diagonal time arguments appear inside the collision integral Eq.$~$\eqref{collision_left}. To close the equation, we use the so-called reconstruction equations$~$\cite{Lipavsky86}
\begin{equation}\label{reconstruction_1}
\begin{split}
{G}^{<}_{}(t,t')=-{G}^{R}_{}(t,t')\rho_{}(t')+\rho_{}(t){G}^{A}_{}(t,t')\dots,\\
\end{split}
\end{equation}
\begin{equation}\label{reconstruction_2}
\begin{split}
&{G}^{>}_{}(t,t')={G}^{R}_{}(t,t')(1-\rho_{}(t'))-(1-\rho_{}(t)){G}^{A}_{}(t,t')+\dots.
\end{split}
\end{equation}
where the dots stand for terms of higher order in $\rho$ and ${G}^{R/A}_{}$. The latter are obtained from Eq.$~$\eqref{retarded_left} and Eq.$~$\eqref{retarded_right}. The sums of all terms of expansions in Eq.$~$\eqref{reconstruction_1} and Eq.$~$\eqref{reconstruction_2} give back the solution of the full KBE by Eqs.$~$\eqref{kbeeq_left}-\eqref{retarded_right}$~$\cite{Lipavsky86}.  In that case, the Greens function components satisfy the spectral identity Eq.\eqref{spectral_identity} and the density matrix $\rho$ from the transport equation connects to the Greens function from the KBE via $\rho(t)=-iG^{<}(t,t)$. We note that there is no practical advantage to formulate the full KBE, Eqs.$~$\eqref{kbeeq_left}-\eqref{retarded_right}, in the alternative way of  Eqs.$~$\eqref{transport}-\eqref{reconstruction_2}, Eq.$~$\eqref{retarded_left} and Eq.$~$\eqref{retarded_right}. The reformulation serves merely as a starting point to introduce the GKBA, as originally done in Ref.$~$\cite{Lipavsky86}. 
\section{Generalized Kadanoff--Baym Ansatz}
The original idea of Ref.$~$\cite{Lipavsky86} is to derive the time-diagonal transport equation for the one-particle density matrix as a generalization of the Boltzmann equation for the quasiparticle occupations. The latter is derivable with the use of the Kadanoff--Baym Ansatz$~$\cite{Kadanoff62}. To derive the time-diagonal transport equation for the one-particle density matrix, the Kadanoff--Baym Ansatz was generalized in Ref.$~$\cite{Lipavsky86}. To achieve this goal, the reconstruction equations Eq.$~$\eqref{reconstruction_1} and Eq.$~$\eqref{reconstruction_2} are truncated at the lowest order, leading to 
\begin{equation}
\begin{split}\label{gkba}
&\tilde{G}^{<}(t,t')=-\tilde{G}^{R}(t,t')\rho(t')+\rho(t)\tilde{G}^{A}(t,t'),\\
&\tilde{G}^{>}(t,t')=+\tilde{G}^{R}(t,t')\bar{\rho}(t')-\bar{\rho}(t)\tilde{G}^{A}(t,t').
\end{split}
\end{equation}
Here, $\bar{\rho}(t)=1-\rho(t)$. The Ansatz in Eq.$~$\eqref{gkba} is known as the Generalized Kadanoff--Baym Ansatz$~$\cite{Lipavsky86}. We note that we have denoted the truncated Greens functions with the tilde symbol on purpose, which is explained below. The tilde function is referred to as the auxiliary Greens function $\tilde{G}$ and is considered to be different from ${G}$, to which we do not have access in the GKBA. The density matrix $\rho(t)$, which appears in Eq.$~$\eqref{gkba}, is obtained from the transport equation Eq.$~$\eqref{transport}, where the auxiliary functions are used in the collision integral, i.e.
\begin{equation}\label{collision_gkba}
\begin{split}
I^{<}(t,t)=\int_{-\infty}^{\infty}d\bar{t}\Sigma^{<}_{}(t,\bar{t})\tilde{G}^{A}_{}(\bar{t},t)+\Sigma^{R}_{}(t,\bar{t})\tilde{G}^{<}_{}(\bar{t},t).
\end{split}
\end{equation}
We note that the selfenergies in Eq.$~$\eqref{collision_gkba} also depend on the auxiliary Greens functions, i.e., $\Sigma^{\lessgtr}_{}[\tilde{G}^{}]$ and $\Sigma^{R}_{}[\tilde{G}^{}]$. It is worth realizing that the auxiliary Greens functions, as a result of their definition in Eq.$~$\eqref{gkba}, automatically satisfy an analog of the spectral identity in Eq.$~$\eqref{spectral_identity}
\begin{equation}
\begin{split}\label{spectral_identity_2}
\tilde{G}^{>}_{}(t,t')-\tilde{G}^{<}_{}(t,t')=\tilde{G}^{R}_{}(t,t')-\tilde{G}^{A}_{}(t,t').
\end{split}
\end{equation}
For the same reason, the relation  $\rho=-i\tilde{G}^{<}$ is also identically satisfied. However, this relation is only a tautology to Eq.$~$\eqref{gkba}.
\section{Hartree-Fock GKBA}
The truncation of the reconstruction equation at first order, i.e., the introduction of the GKBA in  Eq.$~$\eqref{gkba}, induces a non-trivial question. What approximation should we choose in the equations of motion for the auxiliary functions $\tilde{G}^{R}_{}$ and $\tilde{G}^{A}_{}$? On one hand, there is no definitive answer to this question, i.e., the choice is arbitrary. On the other hand, it gives us an opportunity to choose an independent approximation for the auxiliary functions, which can lead to a speedup of the calculations.

Using this freedom, an independent Dyson equation can be assigned to the auxiliary function for $\tilde{G}^{R}_{}$ 
\begin{equation}
\begin{split}\label{retarded_left_aux}
\biggl[i\partial_t-h_{\rm HF}(t)\biggr]\tilde{G}^{R}(t,t')=
\delta(t,t')+\int_{}d\bar{t}\tilde{\Sigma}^{R}(t,\bar{t})\tilde{G}^{R}(\bar{t},t').
\end{split}
\end{equation}
and similar for  $\tilde{G}^{A}_{}$. The level of approximation of $\tilde{G}$ is given by an auxiliary self-energy $\tilde{\Sigma}$ that is different from ${\Sigma}$. The latter expresses the freedom of choice of the independent approximation$~$\cite{Lipavsky86,Hermanns12}. In most practical implementations, the auxiliary self-energy is different from the self-energy in the collision integral $\Sigma\neq\tilde{\Sigma}$. 

It was recognized that the numerical cost can be significantly reduced if we consider the mean-field approximation for the propagation of $\tilde{G}^{R}_{}$, i.e., when the auxiliary selfenergy $\tilde{\Sigma}^{R}_{}(t,\bar{t})=\tilde{\Sigma}^{R}_{}(t)\delta(t-\bar{t})$ is local in time. In closed systems, the popular choice is $\tilde{\Sigma}^{R}_{}(t,\bar{t})=0$. This choice is referred to as the HF-GKBA approximation\cite{Hermanns12}. Here, the auxiliary retarded and advanced Greens functions can be constructed directly from the density matrix, which leads to the dramatic speed-up\cite{Hermanns12,Joost20,Schlunzen20}. It is thus worth to better understand the spectral properties of this approximation. To this end, we first consider the extension of the HF-GKBA into the full two-time domain\cite{Hopjan18}.
\section{Extended Generalized Kadanoff Baym Ansatz}
The GKBA from Eq.$~$\eqref{gkba} allows for the closure of time-diagonal transport equations in Eq.$~$\eqref{transport}. The crucial part is to use Eq.$~$\eqref{gkba} in the equal-time collision integral Eq.$~$\eqref{collision_gkba}. In the scheme known as the eGKBA$~$\cite{Hopjan18}, the same GKBA from Eq.$~$\eqref{gkba} is used to approximate the collision integral of the KBE in Eq.$~$\eqref{collision_left}, i.e., $I[G]\rightarrow I[\tilde{G}]$, which results in 
\begin{equation}\label{collision_egkba}
\begin{split}
I^{<}(t,t')=\int_{-\infty}^{\infty}d\bar{t}\Sigma^{<}_{}(t,\bar{t})\tilde{G}^{A}_{}(\bar{t},t')+\Sigma^{R}_{}(t,\bar{t})\tilde{G}^{<}_{}(\bar{t},t').
\end{split}
\end{equation}
Then, e.g. Eq.$~$\eqref{kbeeq_left} reads as
\begin{equation}
\begin{split}\label{ext1}
\biggl[i\partial_{t}-h^{}_{HF}[\rho](t)\biggr]G^{\lessgtr}_{}(t,t')=\\
\int d\bar{t} [\Sigma^{\lessgtr}_{}[\tilde{G}^{}](t,\bar{t})\tilde{G}^{A}_{}(\bar{t},t')+\Sigma^{R}_{}[\tilde{G}^{}](t,\bar{t})\tilde{G}^{\lessgtr}_{}(\bar{t},t')],
\end{split}
\end{equation}
and similarly for Eq.$~$\eqref{kbeeq_right}. To express precisely the meaning of the eGKBA, in Eq.$~$\eqref{ext1}, we explicitly write the selfenergy dependence on the auxiliary Greens functions. It was shown in Ref.$~$\cite{Hopjan18} that the accompanying equation of motion of the retarded Greens function reads as
\begin{equation}
\begin{split}\label{ext2}
\biggl[i\partial_{t}-h^{}_{HF}[\rho](t)\biggr]G^{R}_{}(t,t')=\\
\delta(t,t')+
\int d\bar{t} \Sigma^{R}_{}[\tilde{G}^{}](t,\bar{t})\tilde{G}^{R}_{}(\bar{t},t'),
\end{split}
\end{equation}
and similarly for ${G}^{A}_{}$. 

Few remarks are in order. First, Eq.$~$\eqref{ext2} is different from the auxiliary Eq.\eqref{retarded_left_aux}, i.e., it contains $\Sigma$ instead of $\tilde{\Sigma}$. Further, in Ref.$~$\cite{Hopjan18}, it was argued that $G^{\lessgtr}_{}(t,t')$ from Eq.$~$\eqref{ext1}, $G^{R}_{}(t,t')$ from Eq.$~$\eqref{ext2}, and $G^{A}_{}(t,t')$ from the corresponding equation are consistent with the spectral identity Eq.$~$\eqref{spectral_identity}. Finally, Eq.$~$\eqref{ext1} reduces to the transport equation Eq.\eqref{transport} on the time diagonal. 

In the eGKBA, the single-particle density matrix is constructed from $G^{<}(t,t')$, which is the solution of Eq.\eqref{ext1}, as
\begin{equation}
\begin{split}\label{rho_gless}
\rho(t)=-iG^{<}(t,t).
\end{split}
\end{equation}
This solution is equal to the GKBA solution $\rho(t)$ obtained from the transport equation in Eq.\eqref{transport}. 
\section{Hartree Fock eGKBA}
As it was the case for the GKBA, the approximation of $\tilde{\Sigma}^{R}$ in the eGKBA can be further specified independently from ${\Sigma}^{R}$. If we use the same approximation for $\tilde{G}^{R}$ as in the HF-GKBA, i.e., $\tilde{\Sigma}^{R}=0$, the whole set of equations is dubbed HF-eGKBA. 

The set of equations is solved as follows. First, $\rho$ is used to construct $\tilde{G}^{R/A}$ via Eq.$~$\eqref{retarded_left_aux}. Then $\rho$ and $\tilde{G}^{R/A}$ are plugged into Eq.$~$\eqref{gkba} and the components $\tilde{G}^{\lessgtr}$ are obtained. Finally, both $\tilde{G}^{\lessgtr}$ and $\tilde{G}^{R/A}$ are used to construct the selfenergy $\Sigma$ on the right hand sides of Eq.$~$\eqref{ext1} and Eq.$~$\eqref{ext2}, and the solutions for ${G}^{\lessgtr}$ and ${G}^{R/A}$ are found. The loop is closed by finding $\rho$ via Eq.$~$\eqref{rho_gless}. 

On one hand, the cheap numerical costs are lost because of the double-time structure of Eq.$~$\eqref{ext1} and Eq.$~$\eqref{ext2} instead of the single-time structure of Eq.$~$\eqref{transport}. On the other hand, the hidden spectral properties of HF-GKBA are revealed in the HF-eGKBA in $G^{\lessgtr}$ and $G^{R/A}$.

\section{Spectral function for homogeneous equilibrium systems} 

To understand the spectral properties of the HF-eGKBA, it is instructive to compare the spectral functions from the HF-eGKBA to the spectral functions from the KBE when the same selfenergy $\Sigma$ is used in both schemes. 

We focus on the equilibrium spectral functions of a homogeneous system. As the system is in equilibrium, we consider the frequency space $\omega$. Similarly, as the system is homogeneous and to avoid possible issues with the non-physical symmetry breaking$~$\cite{Joost21}, the equations are worked out in the basis of the translationally invariant states, i.e., states denoted by quasi-momentum number $k$. 

In non-equilibrium, ${G}^{<}$ and ${G}^{R}$ are independent functions; however, in equilibrium, they are connected by the fluctuation-dissipation relation
\begin{equation}\label{fdr}
G^{<}(\omega,k) =-2 i f(\omega-\mu) {\rm Im}G^{R}(\omega,k),
\end{equation}
where $f(\omega-\mu)$ is the Fermi function and $\mu$ is chosen such that ${\rm Tr}[\rho]=N$, where $N$ is the particle number. Thus, it is sufficient to consider the equation for $G^{R}(\omega,k)$ only. The one-particle density matrix is obtained via
\begin{equation}\label{rho}
\rho(k)=\int-iG^{<}(\omega,k) \frac{d \omega}{2\pi}=-\int f(\omega-\mu) {\rm Im}G^{R}(\omega,k) \frac{d \omega}{\pi},
\end{equation}
and the spectral function as 
\begin{equation}\label{sp}
A(\omega,k) = -\frac{1}{\pi}G^{R}(\omega,k).
\end{equation}
Equations$~$\eqref{fdr}-\eqref{sp} are common to both the HF-eGKBA and the KBE.
\section{Numerical example for Hubbard ring} 

The solution of the KBE and HF-eGKBA spectral functions of the equilibrium (ground) state is illustrated on a homogeneous lattice system of size $L$ with a fixed particle density $n=N/L$, where $N$ is the particle number with equal spin. Concretely, we use a Hubbard ring of $L$ sites  with periodic boundary conditions, which can be expressed as
\begin{equation}
{H}_{\rm ring}=-t\sum_{\langle ij \rangle}^{L}\sum_{\sigma=\uparrow,\downarrow}^{}(c_{i\sigma}^{\dagger}c_{j\sigma}^{}+c.c.)+{U} \sum_{i}^{L}n_{i\uparrow}n_{i\downarrow}
\label{ham}
\end{equation}
where $t$ is the hopping strength and $U$ is the onsite interaction of electrons with opposite spin projections at the same site. Here we consider the case of a half-filled lattice, where the number of electrons for both projections is $N_\uparrow=N_\downarrow=L/2$.
\section{KBE solution} 
As argued above, it is sufficient to consider only the equation of motion for $G^{R}$ in Eq.$~$\eqref{retarded_left}. In equilibrium, we transform it into frequency and quasimomentum space
\begin{equation}\label{retarded_eq1}
\begin{split}
{(G^{R})}^{-1}(\omega,k)={(G^{R}_{HF})}^{-1}(\omega,k)-\Sigma^{R}(\omega,k).
\end{split}
\end{equation}
Writing the Hartree-Fock self-energy explicitly, the equation can be written as
\begin{equation}\label{retarded_eq2}
\begin{split}
{(G^{R})}^{-1}(\omega,k)=\omega-\epsilon_k-U\rho(k)-i\eta-\Sigma^{R}(\omega,k),
\end{split}
\end{equation}
where $\epsilon_k$ is the non-interacting energy dispersion, $\epsilon_k=2t \cos(k)$ and $\eta$
serve as the artificial broadening that helps to numerically converge the equations. As a concrete example of the self-energy $\Sigma^{R}$, we consider the Second Born approximation
\begin{equation}\label{selfen1}
\begin{split}
{\Sigma}^{R}(\omega,k)=\frac{U^{2}}{(2\pi)^{2}}\int d \omega^{'} \int d \omega^{''} \sum_{k^{'}} \sum_{k^{''}} \\
\bigl[ G^{R}_{}(\omega^{'},k^{'})G^{>}_{}(\omega^{''},k^{''})G^{<}_{}(\omega-\omega^{'}+\omega^{''},k^{}-k^{'}+k^{''})+\\
+G^{<}_{}(\omega^{'},k^{'})G^{>}_{}(\omega^{''},k^{''})G^{A}_{}(\omega-\omega^{'}+\omega^{''},k^{}-k^{'}+k^{''})+\\
+G^{<}_{}(\omega^{'},k^{'})G^{R}_{}(\omega^{''},k^{''})G^{<}_{}(\omega-\omega^{'}+\omega^{''},k^{}-k^{'}+k^{''})\bigr]
\end{split}
\end{equation}
Eqs.$~$\eqref{retarded_eq2}-\eqref{selfen1} are iterated together with Eqs.$~$\eqref{fdr}-\eqref{rho} until the convergence of $G^{R}_{}$, which is then used in Eq.$~$\eqref{sp} to construct the spectral function.

\section{HF-eGKBA solution} 
We first transform Eq.$~$\eqref{retarded_left_aux} to frequency and quasimomentum space, and then apply the HF approximation for $\tilde{G}$, i.e., $\tilde{\Sigma}=0$, which leads to
\begin{equation}\label{retarded_aux2}
\begin{split}
{(\tilde{G}^{R})}^{-1}(\omega,k)={(G^{R}_{HF})}^{-1}(\omega,k)=\omega-\epsilon_k-U\rho(k)-i\eta.
\end{split}
\end{equation}
We plug $\tilde{G}^{R}$ and the advanced counterpart $\tilde{G}^{A}$ into Eq.$~$\eqref{gkba} and express them in frequency and quasimomentum space
\begin{equation}
\begin{split}\label{gkba_eq}
&\tilde{G}^{<}_{}(\omega,k)=-G^{R}_{HF} (\omega,k) \rho_{}(k)+\rho_{}(k)G^{A}_{HF}(\omega,k),\\
&\tilde{G}^{>}_{}(\omega,k)=G^{R}_{HF} (\omega,k) \bar{\rho}(k)-\bar{\rho}(k)G^{A}_{HF}(\omega,k).
\end{split}
\end{equation}
Then, $\tilde{G}^{R/A}$ and $\tilde{G}^{\lessgtr}$ are used to construct the Second Born self-energy ${\Sigma}$, explicitly written as
\begin{equation}\label{selfen2}
\begin{split}
{\Sigma}^{R}(\omega,k)=\frac{U^{2}}{(2\pi)^{2}}\int d \omega^{'} \int d \omega^{''} \sum_{k^{'}} \sum_{k^{''}} \\
\bigl[ \tilde{G}^{R}_{}(\omega^{'},k^{'})\tilde{G}^{>}_{}(\omega^{''},k^{''})\tilde{G}^{<}_{}(\omega-\omega^{'}+\omega^{''},k^{}-k^{'}+k^{''})+\\
+\tilde{G}^{<}_{}(\omega^{'},k^{'})\tilde{G}^{>}_{}(\omega^{''},k^{''})\tilde{G}^{A}_{}(\omega-\omega^{'}+\omega^{''},k^{}-k^{'}+k^{''})+\\
+\tilde{G}^{<}_{}(\omega^{'},k^{'})\tilde{G}^{R}_{}(\omega^{''},k^{''})\tilde{G}^{<}_{}(\omega-\omega^{'}+\omega^{''},k^{}-k^{'}+k^{''})\bigr].
\end{split}
\end{equation}
 The equation for $G^{R}$ with $\tilde{G}^{R}=G^{R}_{HF}$ reduces to
\begin{equation}\label{gkbagra_aux}
G^{R}(\omega,k)=G^{R}_{HF}(\omega,k) + G^{R}_{HF}(\omega,k)\Sigma^{R}(\omega,k) G^{R}_{HF}(\omega,k)
\end{equation}
where the selfenergy $\Sigma$ is given by Eq.$~$\eqref{selfen2}. We note that this equation does not have the usual structure of the Dyson equation$~$\cite{Hopjan18}. Eqs.$~$\eqref{retarded_aux2}-\eqref{gkbagra_aux} are iterated together with Eqs.$~$\eqref{fdr}-\eqref{rho} until convergence
of $G^{R}_{}$, which is then used in Eq.$~$\eqref{sp} to construct the spectral function.


\begin{figure}[t!]
\begin{center}
\includegraphics[width=8cm]{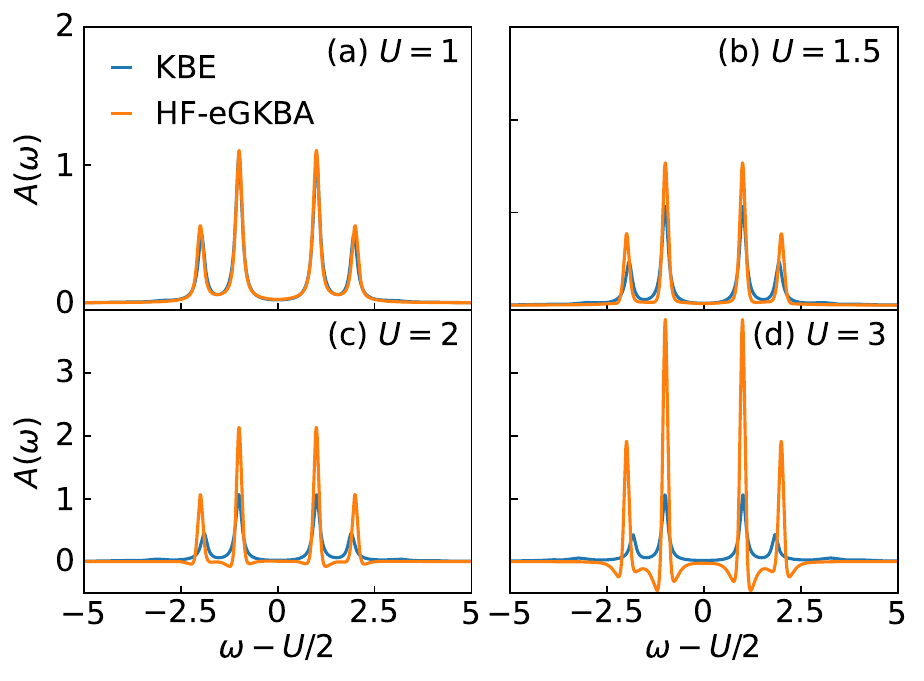}
\caption{The spectral function of homogeneous 6-site Hubbard ring, comparison of the full KBE (blue) and the HF-eGKBA (orange) spectral functions for various interaction $U=1,1.5,2,3$. To achieve the convergence, the 
spectral function are computed with the artificial broadening $\gamma=0.5$. } 
\label{result}
\end{center}
\vspace{-0.8cm}
\end{figure}

\section{Numerical results} 

The KBE and HF-eGKBA spectral functions for the 6-site Hubbard ring are shown in Fig.$~$\ref{result} for various values of the interaction strength $U$. For small interaction strengths $U=1$, the KBE and the HF-eGKBA spectral functions share very similar profiles; see Fig.$~$\ref{result}(a). The difference between the KBE and the HF-eGKBA spectral functions becomes visible for $U=1.5$ in Fig.$~$\ref{result}(b), where the peaks of the HF-eGKBA spectral function become narrower compared to those of the KBE spectral function. Increasing the interaction strength further to $U=2$; see Fig.$~$\ref{result}(c), the HF-eGKBA spectral function still exhibits peaks with similar positions as those of the KBE spectral function. However, most strikingly, the HF-eGKBA spectral function also becomes negative in the vicinity of the main peaks. For even higher interaction strengths $U=3$, see Fig.$~$\ref{result}(d), the widths of the regions of the negative HF-eGKBA spectral function further grow, and the negative values become more pronounced. 

\section{Discussion} 

The emergence of the negative spectral function in HF-eGKBA constitutes the main result of the present work. This opens up further scenarios. With the negative spectral function at hand, it is possible that, for inhomogeneous systems, the integral of the lesser function in Eq.$~$\eqref{rho} will produce negative entries on the diagonal of $\rho$, i.e., the negative densities $n_i=\rho_{ii}$. This feature is clearly non-physical.

Moreover, the possibility of negative densities can proliferate to the standard non-equilibrium time-dependent HF-GKBA. Indeed, the negative values of the densities $n_i=\rho_{ii}$ were already observed in the time propagation of HF-GKBA equations of electrons\cite{Schlunzen20}, and also in the bosonic systems with mean-field GKBA equations\cite{Pavlyukh22b}. Additionally, the negative spectral features hidden in the collision integral are likely the reason for the instabilities of the HF-GKBA equations\cite{Schlunzen20}.

As conjectured in Ref.$~$\cite{Hopjan18}, the main source of negative spectral weights is the structure of Eq.$~$\eqref{gkbagra_aux}, i.e., $G=G_{HF}+G_{HF}\Sigma G_{HF}$, which is not the standard structure of the Dyson equation. We note that in Ref.$~$\cite{Hopjan18}, we bypassed this problem by replacing Eq.$~$\eqref{gkbagra_aux} with an equation that has the usual Dyson structure, i.e., $G=G_{HF}+G_{HF}\Sigma G_{}$. Doing so for the present model, the negative spectral functions were also not observed (not shown). 

Thus, in the standard HF-GKBA, the appearance of the negative densities originates from the structure of the collision integral; see Eq.$~$\eqref{collision_gkba}. We note that this structure is crucial for the linear-in-time scaling of the computational costs, as in the G$_1$-G$_2$ scheme$~$\cite{Joost20,Schlunzen20}. Thus, the possibility of the emergence of negative densities and instabilities for larger interaction strength is a trade off for the gained computational speed-up. That being said, for lower interaction, these issues do not appear to be so severe. We note that a similarly good comparison between KBE and GKBA has been found in the transport setup$~$\cite{Cosco20}. It would be interesting to test whether the situation would improve for the recently proposed extensions of the GKBA$~$\cite{Kalvova23,Kalvova24,Kalvova25,Pavlyukh25}.

\acknowledgments
We acknowledge support from the Polish National Agency for Academic Exchange (NAWA)’s Ulam Programme (project BNI/ULM/2024/1/00124).


\end{document}